\documentclass[final,5p,times,twocolumn,number]{elsarticle}

\usepackage{amssymb,amsmath}
\usepackage{graphicx}
\graphicspath{{figs/}}
\usepackage{booktabs}
\usepackage{array}
\usepackage{multirow}
\usepackage{caption}
\usepackage{subcaption}
\usepackage{float}
\usepackage{xcolor}
\usepackage{svg}
\usepackage{enumitem}
\usepackage{mdframed}
\usepackage{longtable}
\usepackage{booktabs}
\usepackage{tikz}
\usetikzlibrary{positioning,arrows.meta,shapes.geometric,fit,backgrounds,shadows}
\usepackage[hidelinks,colorlinks=true,
            linkcolor=blue,citecolor=blue,urlcolor=blue]{hyperref}

\definecolor{datacolor}{RGB}{0,100,160}
\definecolor{modelcolor}{RGB}{20,120,60}
\definecolor{explaincolor}{RGB}{160,60,0}
\definecolor{evalcolor}{RGB}{120,0,120}
\definecolor{statsbg}{RGB}{22,22,34}
\definecolor{statstext}{RGB}{220,220,220}
\definecolor{statsaccent}{RGB}{160,110,230}
\definecolor{aagrey}{RGB}{90,90,90}

\mdfdefinestyle{statsbox}{%
  backgroundcolor=statsbg, linecolor=statsaccent, linewidth=1.2pt,
  roundcorner=4pt, innerleftmargin=8pt, innerrightmargin=8pt,
  innertopmargin=6pt, innerbottommargin=6pt, skipabove=6pt, skipbelow=4pt}

\tikzset{
  databox/.style={rectangle,draw=datacolor,thick,fill=datacolor!8,
    text width=3.3cm,align=center,rounded corners=3pt,
    minimum height=0.75cm,font=\scriptsize,drop shadow},
  modelbox/.style={rectangle,draw=modelcolor,thick,fill=modelcolor!8,
    text width=3.3cm,align=center,rounded corners=3pt,
    minimum height=0.75cm,font=\scriptsize,drop shadow},
  explainbox/.style={rectangle,draw=explaincolor,thick,fill=explaincolor!8,
    text width=3.3cm,align=center,rounded corners=3pt,
    minimum height=0.75cm,font=\scriptsize,drop shadow},
  evalbox/.style={rectangle,draw=evalcolor,thick,fill=evalcolor!8,
    text width=11.0cm,align=center,rounded corners=3pt,
    minimum height=0.80cm,font=\scriptsize,drop shadow},
  colhead/.style={font=\small\bfseries,inner sep=3pt},
  arrow/.style={-{Stealth[length=5pt]},thick},
  xarrow/.style={-{Stealth[length=5pt]},thick,dashed,draw=aagrey},
}

\journal{Astronomy \& Computing}

\begin{document}
\begin{frontmatter}

\title{CASPER: Interpretable ResNet based Classifier with FastShap Explainer for Gravitational Wave Detection  }

\author[1]{R.Rai\corref{cor1}}
\ead{rashika.rai2004@gmail.com }
\author[2]{R.Verma}
\author[3]{Somya}
\cortext[cor1]{Corresponding author}
\affiliation[1]{organization={Department of Physics, Shaheed Rajguru College of Applied Sciences for Women}, city={University of Delhi}, country={PIN 110096 India }}
\affiliation[2]{organization={Assistant Professor,Department of Physics, Shaheed Rajguru College of Applied Sciences for Women}, city={University of Delhi}, country={PIN 110096 India }}
\affiliation[3]{organization={Department of Physics, Shaheed Rajguru College of Applied Sciences for Women}, city={University of Delhi}, country={PIN 110096 India }}

\begin{abstract}
Traditional matched filtering has been the standard for Gravitational waves (GW) detection ever since LIGO was established, even though it requires pre-computed waveform templates and provides no accounts of information about which signal drove the decision of classification. Deep-learning alternatives showed competitive sensitivity, but system biases—including class overlap, imbalanced class weighting, limited sample variation, and train–test mismatch—continue to cause problems with generalisation in real detector noise. We introduce CASPER-Classification with Attribution via ShaPlEy in Residual neural networks, an end-to-end pipeline combining residual convolutional neural network (CNN) classifier with a FastSHAP explainer. 260  distinct events from the Gravitational Wave open Science Centre were fetched across SNR range of 7-42 from both H1 and L1 detectors with no synthetic augmentation. The classifier achieves AUC (Area Under Curve) of 91\% across the model with a low false alarm rate. Focal Loss and Platt Calibration were used to improve decision boundary and generalisation. FastSHAP attribution maps recover the complete chirp morphology and provides detailed maps for a visual interpretation of the decision. The complete pipeline contains fewer parameters than standard deep learning models and requires no hardware except a standard CPU making our model an effective lightweight pipeline for Gravitational Wave Detection under real life conditions.
\end{abstract}

\begin{keyword}
gravitational waves \sep methods: data analysis \sep neural networks \sep software: public \sep stars: neutron \sep LIGO
\end{keyword}
\end{frontmatter}

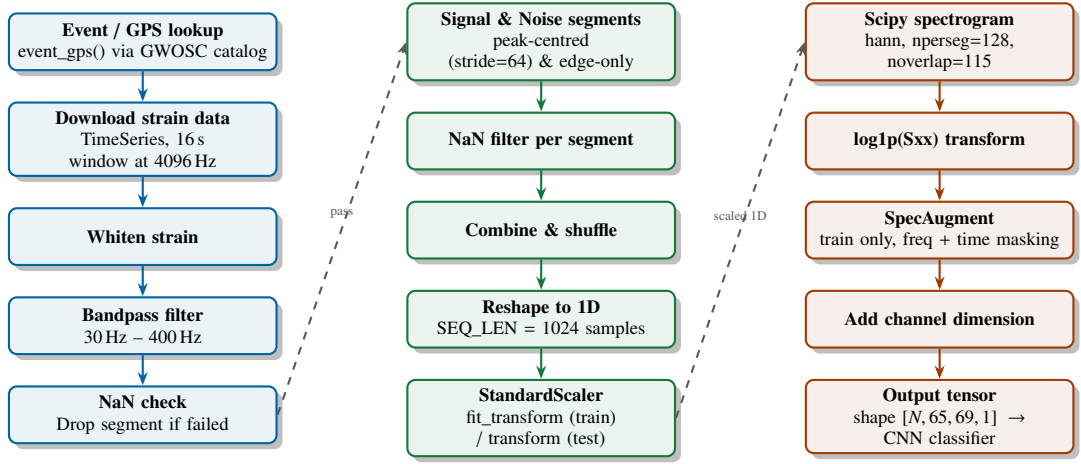
\begin{figure*}[t]
  \centering
  \begin{tikzpicture}[node distance=0.40cm and 1.7cm]
    \node[databox] (d1) {\textbf{Event / GPS lookup}\\event\_gps() via GWOSC catalog};
    \node[databox,below=of d1] (d2) {\textbf{Download strain data}\\TimeSeries, 16\,s window at 4096\,Hz};
    \node[databox,below=of d2] (d3) {\textbf{Whiten strain}};
    \node[databox,below=of d3] (d4) {\textbf{Bandpass filter}\\30\,Hz -- 400\,Hz};
    \node[databox,below=of d4] (d5) {\textbf{NaN check}\\Drop segment if failed};

    \node[modelbox,right=of d1] (m1) {\textbf{Signal \& Noise segments}\\peak-centred (stride=64) \& edge-only};
    \node[modelbox,below=of m1] (m2) {\textbf{NaN filter per segment}};
    \node[modelbox,below=of m2] (m3) {\textbf{Combine \& shuffle}};
    \node[modelbox,below=of m3] (m4) {\textbf{Reshape to 1D}\\SEQ\_LEN = 1024 samples};
    \node[modelbox,below=of m4] (m5) {\textbf{StandardScaler}\\fit\_transform (train) / transform (test)};

    \node[explainbox,right=of m1] (e1) {\textbf{Scipy spectrogram}\\hann, nperseg=128, noverlap=115};
    \node[explainbox,below=of e1] (e2) {\textbf{log1p(Sxx) transform}};
    \node[explainbox,below=of e2] (e3) {\textbf{SpecAugment}\\train only, freq + time masking};
    \node[explainbox,below=of e3] (e4) {\textbf{Add channel dimension}};
    \node[explainbox,below=of e4] (e5) {\textbf{Output tensor}\\shape $[N, 65, 69, 1] \to$ CNN classifier};

    \node[colhead,text=datacolor,   above=0.35cm of d1]{ACQUISITION \& CONDITIONING};
    \node[colhead,text=modelcolor,  above=0.35cm of m1]{SEGMENTATION \& SCALING};
    \node[colhead,text=explaincolor,above=0.35cm of e1]{SPECTROGRAM GENERATION};

    \foreach \a/\b/\c in {d1/d2/datacolor,d2/d3/datacolor,d3/d4/datacolor,d4/d5/datacolor}
      \draw[arrow,draw=\c] (\a)--(\b);
    \foreach \a/\b/\c in {m1/m2/modelcolor,m2/m3/modelcolor,m3/m4/modelcolor,m4/m5/modelcolor}
      \draw[arrow,draw=\c] (\a)--(\b);
    \foreach \a/\b/\c in {e1/e2/explaincolor,e2/e3/explaincolor,e3/e4/explaincolor,e4/e5/explaincolor}
      \draw[arrow,draw=\c] (\a)--(\b);

    \draw[xarrow] (d5.east)--node[above,font=\tiny,text=aagrey]{pass}(m1.west);
    \draw[xarrow] (m5.east)--node[above,font=\tiny,text=aagrey]{scaled 1D}(e1.west);

  \end{tikzpicture}
  \caption{\textbf{Data Pipeline Architecture.}
    Data cleaning and Spectrogram creation}
  \label{fig:datapipeline}
\end{figure*}

\section{Introduction}
\label{sec:intro}

 The first Gravitational wave (GW) detection back in 2015 using the LIGO  detectors opened a new field of astronomy letting us both hear and see the celestial events\citep{Abbott2016,Abbott2017,Abbott2021,Abbott2023}. The current detection method of using Matched Filtering which utilised pre-computed waveform templates  \citep{Allen2012,Dhurandhar1994,Owen1999} has been proved a massive success and led to the discovery of around 90 Events in the observing runs\citep{Abbott2023}. 
 But matched filtering even though the most accurate method till yet chose to be a liability due to its no feature of knowing why a GW is classified as GW and its computational costs skyrocket due to their increasing sizes because of increase in sensitivity.\\

 Deep learning, right at the time emerged as the perfect alternative to matched filtering, providing better latency and sensitivity with a low false alarm rate\citep{George2018,Gabbard2018,Gebhard2019,Corizzo2020,Yan2022}. But the biggest issue with deep learning was the large number of parameters making it require bigger computational power and had no account of what drove the model to decision, i.e; no way to justify results. In addition to this, the understood study on the following learning biases- train-test batch, class overlap , imbalanced weighting and many more makes them also unreliable for GW detection \citep{Nagarajan2025}. Over the years many models were created to deal with the issues and even though many achieved phenomenal probability, the most common issue was that most models didn't have real data training, had huge number of parameters needing large GPUs or no accounts of what drove the decision of the probability by the model.\\
During the third observing run, we saw current models with nearly perfect algorithms including Aframe pipeline \citep{Marx2023} that created an end to end purely Machine Learning(ML)  based pipeline giving the classifier significant low inference latency of 22ms. But the issue remains same; no interpretability or justification of decision or generalisation of model on Low mass mergers or harder signals.
Models that worked on low mass mergers are; due to huge number of parameters; extremely computationally costly and also compromise with the latency \citep{Beheshtipour2021}. Hence, a clear trade off between accuracy, latency and SNR generalisation along with lack of interpretability remains unsolved.\\
In this paper, we present \textit{CASPER:Classification with Attribution via Shapley in Residual neural networks}. The model presented here is a hybrid of a customised 7 layer ResNet with a 2D CNN. CNNs have always been a trustworthy method for GW detection but merging it with custom ResNet along with no use of synthetic data injections makes it a promising candidate for GW classification at a generalised SNR without compromising on latency that complements existing models with its interpretability and computational efficiency. The paper will discuss the data acquisition and cleaning techniques, a brief overview over CASPER's methodology and the metrics used to assess the performance along with the use of FastSHAP \citep{Jethani2022} as an explainer in the said order.

\section{Methods}
\label{sec:methods}

\subsection{Data Acquisition and Conditioning}
\label{sec:data}
The strain data required for the model training and testing were automatically fetched from the LIGO open science center (GWOSC) site \citep{Vallisneri2015}. The data downloaded was sampled at 4096 Hz at 16s using GWpy library \citep{Macleod2021}. A proper visual of the pipeline can be seen in Figure ~\ref{fig:datapipeline}. The data statistics and calculation of how many spectrograms were created can also be found in Table ~\ref{tab:segconfig} which summarises a thorough explanation of our data division in detailed format.\\
The entire catalogue of GWOSC was filtered for 260 distinct events which were fetched around the merger time with about 80\% data used for training and rest as a test set. To fix the inherent class imbalance of GW data; Spec Augment is used for creating masked frequencies sample without cropping,rotating or damaging the chirp morphology so as to increase the number of signal samples to fix the imbalance. No synthetic injections are made in the data whatsoever. The whole data covers the SNR range of 7-42 and is randomly chosen to simulate a real life detector. A detailed pipeline of cleaning, whitening and band-passing \citep{Butterworth1930} is used for pre-processing of data before being input in the model.

\begin{table}[h]
  \centering
  \caption{Segment configuration and sample counts per event for the
    CASPER pipeline. Values are derived directly from
    \texttt{PipelineConfig} in the pipeline code. The table is divided
    into four blocks by mid-rules: global acquisition parameters,
    positive (signal) segment extraction, noise segment extraction, and
    per-event and corpus totals. All 260 events (194 train, 66 test)
    are drawn by random selection from the GWTC catalogue spanning SNR
    7--42, with no SNR-tier stratification. Theoretical maximums are shown before NaN rejection. After an approximate 6\% drop due to NaN values, the final train and test segments are 27,010 and 7,250, respectively.}
  \label{tab:segconfig}
  \renewcommand{\arraystretch}{1.25}
  \begin{tabular}{@{}lll@{}}
    \toprule
    \textbf{Quantity} & \textbf{Value / Formula} & \textbf{Result} \\
    \midrule
    Data duration        & Centred on GPS time       & 16\,s \\
    Sampling rate        & Standard LIGO high-freq.  & 4096\,Hz \\
    Total samples/16\,s  & $16\times4096$            & 65\,536 \\
    Segment length       & \texttt{SEQ\_LEN}          & 1024 \\
    \midrule
    \multicolumn{3}{@{}l@{}}{\itshape Positive (signal) segments} \\
    Peak span  & $\pm$1024 samples         & 2048\,pts \\
    Stride     & \texttt{STRIDE} = 64 samp.& --- \\
    Segs/det.  & $(2048-1024)/64$          & 16 \\
    \midrule
    \multicolumn{3}{@{}l@{}}{\itshape Noise segments} \\
    Stride       & 1024 (non-overlap)       & --- \\
    Excluded zone& Central third            & --- \\
    Noise segs/det.& Outer thirds           & $\approx$42 \\
    \midrule
    \textbf{Total/event/det.}  & $16+42$            & $\approx$\textbf{58} \\
    Detectors used             & H1\,+\,L1          & 2 \\
    Spectrograms/event         & $58\times2$        & $\approx$116 \\
    \midrule
    Train corpus    & 194\, $\times$116 & $\approx$22,504 \\
    Val split (15\%) & from train        & $\approx$3,376 \\
    Test corpus     & 66\,$\times$116  & $\approx$7,656 \\
    Spectrogram shape& STFT (128, 90\% OL)& $(65,69,1)$ \\
    \midrule
    \multicolumn{3}{@{}l@{}}{\itshape Final Counts (After 6\% NaN Drop)} \\
    Final Train Segments & & \textbf{27,010} \\
    Final Test Segments  & & \textbf{7,250} \\
    \bottomrule
  \end{tabular}
\end{table}
\subsection{Dataset Construction}
\label{sec:dataset}

\subsubsection{Training Dataset}
\label{sec:traindata}
\begin{figure}[h]
  \centering
  \includegraphics[width=\columnwidth]{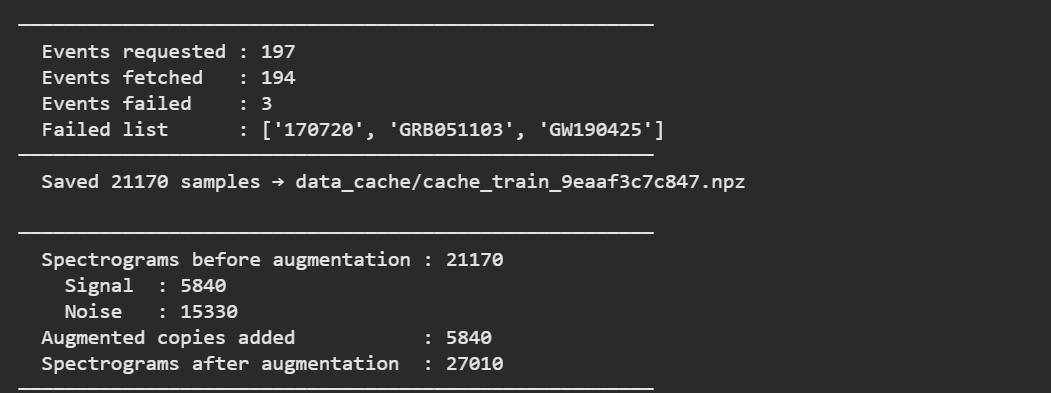}
  \caption{Training dataset compilation output (terminal log).  A total of 194 events were successfully retrieved (3
    failed). After window extraction and SpecAugment augmentation, the training
    set contains 27\,010 spectrograms (5840 raw signal $+$ 5840 augmented
    $+$ 15\,330 noise).}
  \label{fig:screenshot_train}
\end{figure}

Training dataset is made of 194 events fetched directly from the GWOSC site using the GWpy library. These 194 events contributed to 22504 spectrograms of both signal and noise. Around 6\% of the segments were dropped because LIGO fills the missed data with NaN values for the public data. The positive windows were extracted by a 90\% overlap in the neighborhood of the merger GPS time, producing approximately 16 spectrograms of signal per event per detector. The noise samples are extracted from outer thirds with no overlap producing 42 samples per event per detector giving us 58 spectrograms per event per detector. A 15\% of the train set is used for validation to monitor early stopping and Platt Calibration Fitting. Spec Augment is used to add morphological variety to minority class \citep{Park2019}. Figure ~\ref{fig:screenshot_train}  shows the dataset compilation terminal log.

\subsubsection{Test Dataset}
\label{sec:testdata}
\begin{figure}[h]
  \centering
  \includegraphics[width=\columnwidth]{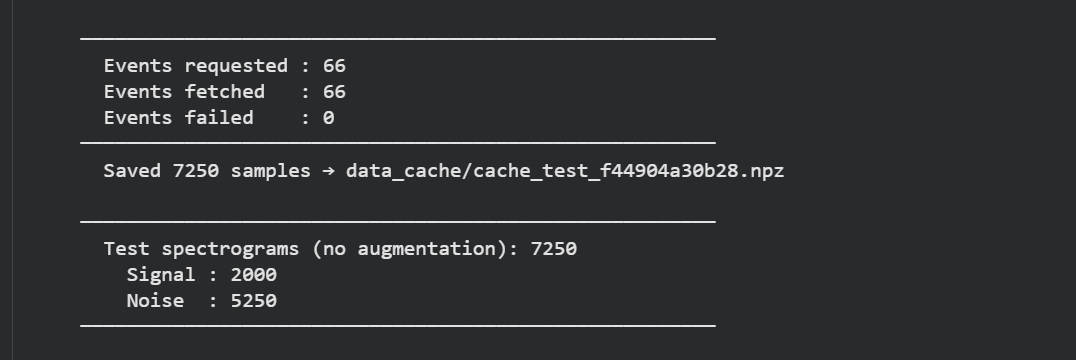}
  \caption{Test dataset compilation output (terminal log).  All 66 requested events
    were successfully retrieved (0 failures). The test set contains 7250
    spectrograms (2000 signal, 5250 noise) with no augmentation applied.}
  \label{fig:screenshot_test}
\end{figure}

The test data is made of 66 unseen events to avoid data leakage spanning the same SNR range. No augmentation was applied to simulate real world class imbalance. The extraction yielded 2000 signal samples and 5250 noise samples making a total of 7250 spectrograms. Figure~\ref{fig:screenshot_test} shows the test set compilation statistics.

\subsubsection{Bias Mitigation in Data Design}
\label{sec:databias}

Supervised GW detection has a number of well-documented failure modes due to the way training data are structured \citep{Gebhard2019,Corizzo2020}. We have addressed these directly with our approach as follows:
Reduced class overlap by limiting signal windows to the direct vicinity of the merger GPS time; this sharpens the morphological class separation compared to the outer third (of the total) noise windows.
Class imbalance was also addressed using class weighted training and focal loss \citep{Lin2017}; the latter will reduce the weight for well classified examples while enhancing the gradient signal at the difficult classification thresholds.
Variability was addressed using SpecAugment signal augmentation on both classes of signals during training; we further had variability of real detector noise based upon 194 different O3a, O3b and O4 observing run events (the latter provide variability in PSD, non-stationary and glitch content among individual segments \citep{Davis2021,Cabero2019}).

\subsection{Spectrogram Representation}
\label{sec:spec}

Each 0.25-second window was normalised with a \texttt{StandardScaler} \citep{Pedregosa2011} fitted on training data only, then converted to an STFT spectrogram (Hann window 128 samples, 90\% overlap). Log-compression $S \rightarrow \log(1 + S)$ yielded spectrograms of shape $(65, 69, 1)$. The 65-bin frequency axis spans 0--2048\,Hz at the Hann window resolution; the 69-frame time axis covers the 0.25-second window at 90\% overlap. Log-compression standardises power contributions across the frequency axis, reducing the tendency of CNNs to disproportionately weight high-amplitude features \citep{He2016}, and compresses the wide dynamic range of GW strain into a representation more tractable for gradient-based optimisation.

\subsection{Pipeline Architecture and Training}
\label{sec:pipeline}

\subsubsection{System Overview}
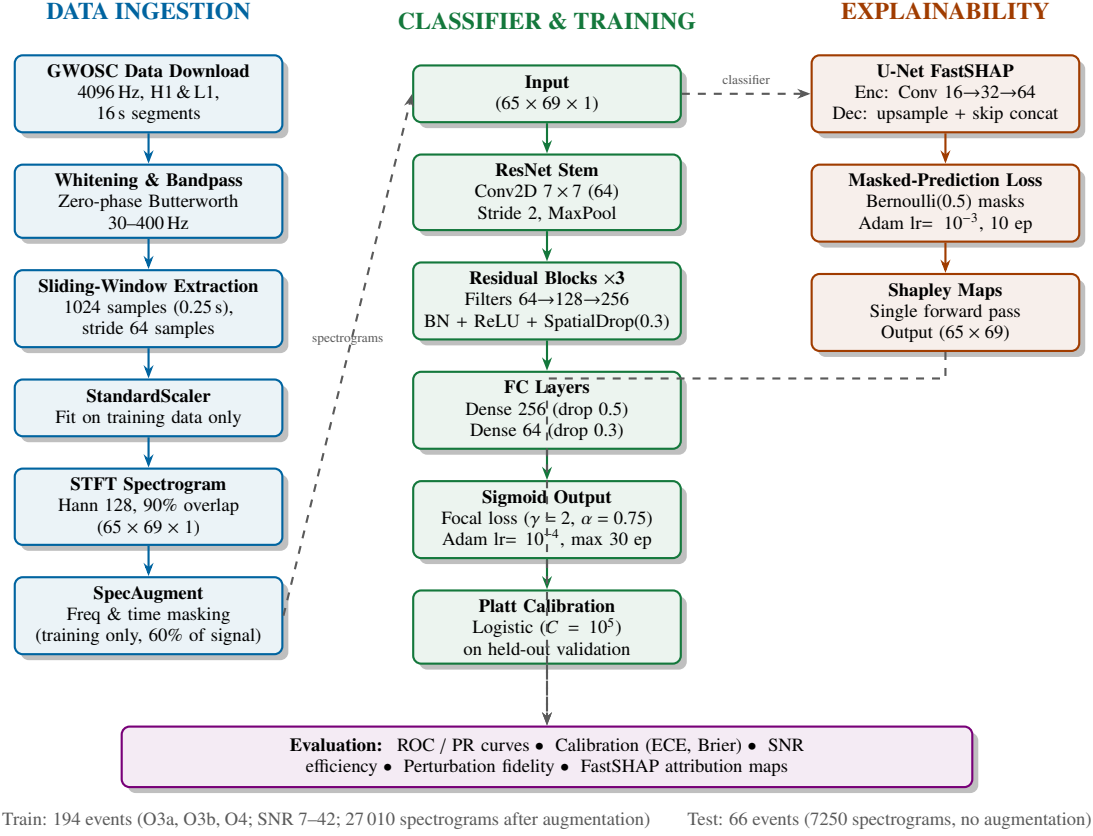
\begin{figure*}[t]
  \centering
  \begin{tikzpicture}[node distance=0.40cm and 1.7cm]
    \node[databox] (d1) {\textbf{GWOSC Data Download}\\4096\,Hz, H1\,\&\,L1, 16\,s segments};
    \node[databox,below=of d1] (d2) {\textbf{Whitening \& Bandpass}\\Zero-phase Butterworth\\30--400\,Hz};
    \node[databox,below=of d2] (d3) {\textbf{Sliding-Window Extraction}\\1024 samples (0.25\,s),\\stride 64 samples};
    \node[databox,below=of d3] (d4) {\textbf{StandardScaler}\\Fit on training data only};
    \node[databox,below=of d4] (d5) {\textbf{STFT Spectrogram}\\Hann 128, 90\% overlap\\$(65\!\times\!69\!\times\!1)$};
    \node[databox,below=of d5] (d6) {\textbf{SpecAugment}\\Freq \& time masking\\(training only, 60\% of signal)};
    \node[modelbox,right=of d1] (m1) {\textbf{Input}\\$(65\!\times\!69\!\times\!1)$};
    \node[modelbox,below=of m1] (m2) {\textbf{ResNet Stem}\\Conv2D $7\!\times\!7$ (64)\\Stride 2, MaxPool};
    \node[modelbox,below=of m2] (m3) {\textbf{Residual Blocks $\times$3}\\Filters 64$\to$128$\to$256\\BN + ReLU + SpatialDrop(0.3)};
    \node[modelbox,below=of m3] (m4) {\textbf{FC Layers}\\Dense 256 (drop 0.5)\\Dense 64 (drop 0.3)};
    \node[modelbox,below=of m4] (m5) {\textbf{Sigmoid Output}\\Focal loss ($\gamma\!=\!2$, $\alpha\!=\!0.75$)\\Adam lr$=10^{-4}$, max 30 ep};
    \node[modelbox,below=of m5] (m6) {\textbf{Platt Calibration}\\Logistic ($C\!=\!10^5$)\\on held-out validation};
    \node[explainbox,right=of m1] (e1) {\textbf{U-Net FastSHAP}\\Enc: Conv 16$\to$32$\to$64\\Dec: upsample + skip concat};
    \node[explainbox,below=of e1] (e2) {\textbf{Masked-Prediction Loss}\\Bernoulli(0.5) masks\\Adam lr$=10^{-3}$, 10 ep};
    \node[explainbox,below=of e2] (e3) {\textbf{Shapley Maps}\\Single forward pass\\Output $(65\!\times\!69)$};
    \node[evalbox,below=2.2cm of m5] (ev)
      {\textbf{Evaluation:}\;
        ROC / PR curves\;\textbullet\;
        Calibration (ECE, Brier)\;\textbullet\;
        SNR efficiency\;\textbullet\;
        Perturbation fidelity\;\textbullet\;
        FastSHAP attribution maps};
    \node[colhead,text=datacolor,   above=0.35cm of d1]{DATA INGESTION};
    \node[colhead,text=modelcolor,  above=0.35cm of m1]{CLASSIFIER \& TRAINING};
    \node[colhead,text=explaincolor,above=0.35cm of e1]{EXPLAINABILITY};
    \foreach \a/\b/\c in {d1/d2/datacolor,d2/d3/datacolor,d3/d4/datacolor,d4/d5/datacolor,d5/d6/datacolor}
      \draw[arrow,draw=\c] (\a)--(\b);
    \foreach \a/\b/\c in {m1/m2/modelcolor,m2/m3/modelcolor,m3/m4/modelcolor,m4/m5/modelcolor,m5/m6/modelcolor}
      \draw[arrow,draw=\c] (\a)--(\b);
    \foreach \a/\b/\c in {e1/e2/explaincolor,e2/e3/explaincolor}
      \draw[arrow,draw=\c] (\a)--(\b);
    \draw[xarrow] (d6.east)--node[above,font=\tiny,text=aagrey]{spectrograms}(m1.west);
    \draw[xarrow] (m1.east)--node[above,font=\tiny,text=aagrey]{classifier}(e1.west);
    \draw[xarrow] (m6.south)--++(0,-0.35)-|(ev.north);
    \draw[xarrow] (e3.south)--++(0,-0.35)-|(ev.north);
    \node[font=\scriptsize,text=aagrey,below=0.2cm of ev]
      {Train: 194 events (O3a, O3b, O4; SNR 7--42; 27\,010 spectrograms after augmentation)\qquad
       Test: 66 events (7250 spectrograms, no augmentation)};
  \end{tikzpicture}
  \caption{\textbf{CASPER pipeline architecture.}
    \textit{Left (blue):} Data ingestion and preprocessing.
    \textit{Centre (green):} ResNet CNN classifier.
    \textit{Right (orange):} U-Net FastSHAP explainer
    (Eq.~\ref{eq:fastshap}).
    \textit{Bottom (purple):} Shared evaluation outputs.}
  \label{fig:architecture}
\end{figure*}

The CASPER pipeline (see Fig.~\ref{fig:architecture}) has three parallel components: a data ingestion module, a Residual CNN classifier, and a FastSHAP explainer. We convert raw strain data from 4096 Hz to $65\times69$ STFT spectrograms. Before that we normalize and whiten the raw 4096\,Hz strain data and filter it between 30 -- 400\,Hz. For increased robustness we also perform time and frequency masking when we train our model using SpecAugment.

\subsubsection{Classifier and Calibration}
We used the ResNet-inspired architecture \citep{He2016} for our CNN classifier. It has a $7\times7$ convolutional stem followed by three residual blocks with filter sizes of 64, 128, and 256. In addition to SpatialDropout2D ($p=0.3$) to prevent overfitting, we also applied $\ell_2$ regularization ($10^{-4}$).

In order to drive training we use Focal Loss \citep{Lin2017} ($\gamma=2$, $\alpha=0.75$). This loss function gives high weight to difficult cases where the signal-to-noise ratio (SNR) is in the 7 -- 10 (Low SNR) range. Optimization was done using Adam ($lr=10^{-4}$) along with early stopping and learning rate scheduling. Finally, for the purpose of evaluating the reliability of our model, we performed Platt calibration \citep{Platt1999} on held out validation data. The results were evaluated based on Expected Calibration Error (ECE) and Brier scores.

\subsubsection{FastSHAP Explainer}
\label{sec:fastshap}
For real-time interpretability, we trained a U-Net \citep{Ronneberger2015} to approximate pixel-wise Shapley values. The architecture uses skip connections to maintain high spatial resolution in the attribution maps. The explainer is optimized via a masked-prediction consistency objective:
\begin{equation}
  \mathcal{L} = \mathbb{E}_{\mathbf{m}}\!\left[\!\left(\sum_j\phi_j(\mathbf{x})m_j
    - \bigl[f(\mathbf{x}\odot\mathbf{m})-f(\mathbf{x}_\mathrm{bg})\bigr]\right)^{\!2}\right],
  \label{eq:fastshap}
\end{equation}

where $\mathbf{m}$ is a random binary mask and $\phi_j$ is the attribution for pixel $j$ which allows single pass generation of attribution maps, providing support for classifier's predictions.
\begin{figure*}[t!]
  \centering
  \includegraphics[width=\textwidth]{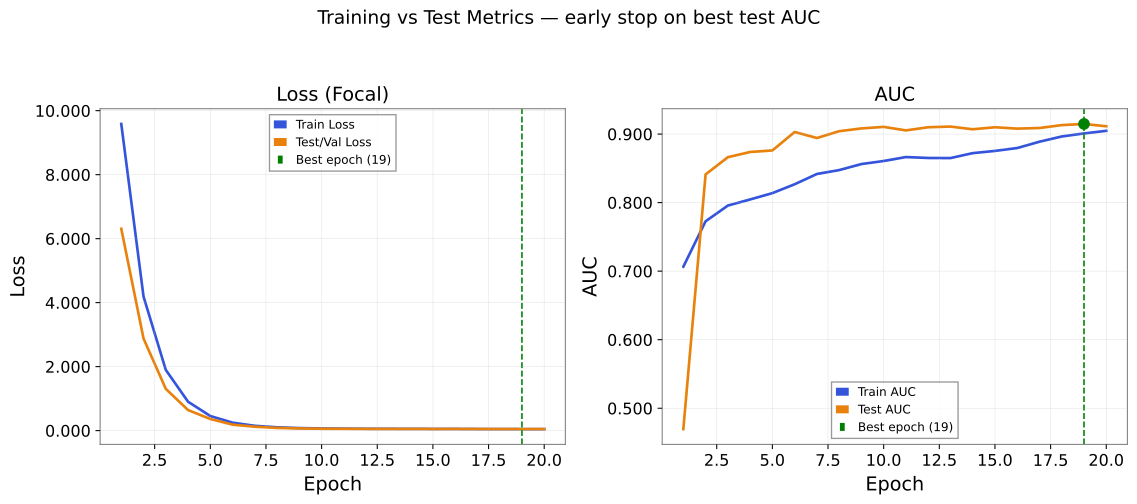}
  \caption{Training history across all epochs. Early stopping (patience 8) terminates training at the epoch of maximum validation AUC, with the best-checkpoint model saved automatically.}
  \label{fig:training_history}
\end{figure*}
\section{Results}
\label{sec:results}

\subsection{Training History}
\label{sec:training_history}

Figure ~\ref{fig:training_history} depicts the model statistics with the training vs test AUC and focal loss. The model trained with 194 events at 20 epochs is subjected to early stopping(patience 8) and ReduceLROnPlateau (patience 4, factor 0.5) on validation set AUC to deal with overfitting. The model achieves the best results at epoch 19. Focal loss and balanced class weights make sure harder segments are also contributing meaningfully. Log Loss is monitored as a metric on the whole model and reported at the end to ensure authenticity of the model.

\subsection{Classification Performance}
\label{sec:classperf}

\begin{figure}[h]
  \centering
  \includegraphics[width=\columnwidth]{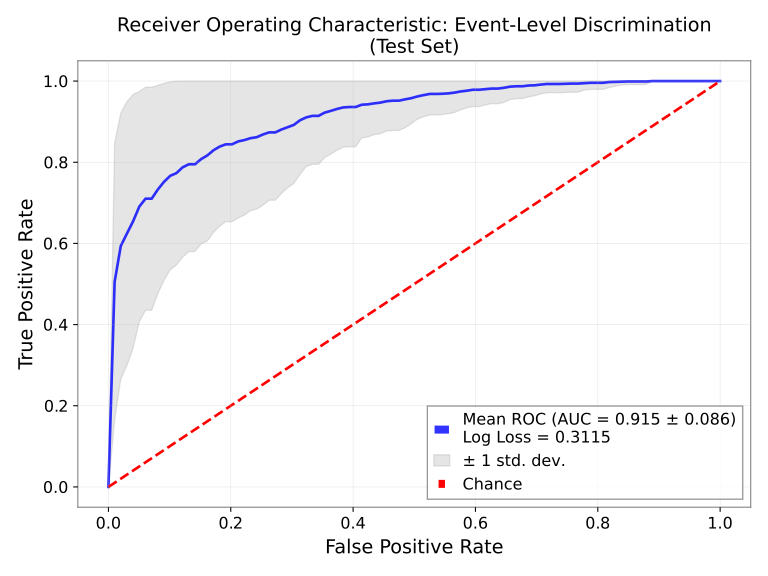}
  \caption{Mean per-event ROC curve ($\pm 1\sigma$ shaded band)  Each event contributes an individual per-event AUC; the mean and
    standard deviation across events are reported in the legend. }
  \label{fig:roc}
\end{figure}

Table ~\ref{tab:metrics} presents the classification report at three thresholds: the default ($\tau = 0.50$), the precision optimised one where the model provides the highest precision along with sufficient recall that is identified by the pipeline itself and a high precision, lowest FAR (False Alarm Rate) threshold for the GW detection. The model achieved a mean AUC per event of 0.9152 in the unbiased threshold along with a mean Log Loss per event of 0.3115 (see figure ~\ref{fig:roc}).

\begin{table}[h]
\centering
\caption{Classification metrics for CASPER.}
\label{tab:metrics}
\renewcommand{\arraystretch}{1.2}
\resizebox{\columnwidth}{!}{%
\begin{tabular}{lccccl} 
\toprule
\textbf{Threshold} & \textbf{Precision} & \textbf{Recall} & \textbf{F1} & \textbf{FPR} & \textbf{Mode} \\
\midrule
0.5000 & 0.8357 & 0.6865 & 0.7538 & 0.0514 & Default \\
0.7927 & 0.9539 & 0.5275 & 0.6793 & 0.0097 & Optimal (High Precision) \\
0.8500 & 0.9677 & 0.4790 & 0.6408 & 0.0061 & High Precision \\
\bottomrule
\end{tabular}%
}
\end{table}

\subsection{Calibration}
\label{sec:calib_res}
\begin{figure}[h]
  \centering
  \includegraphics[width=\columnwidth]{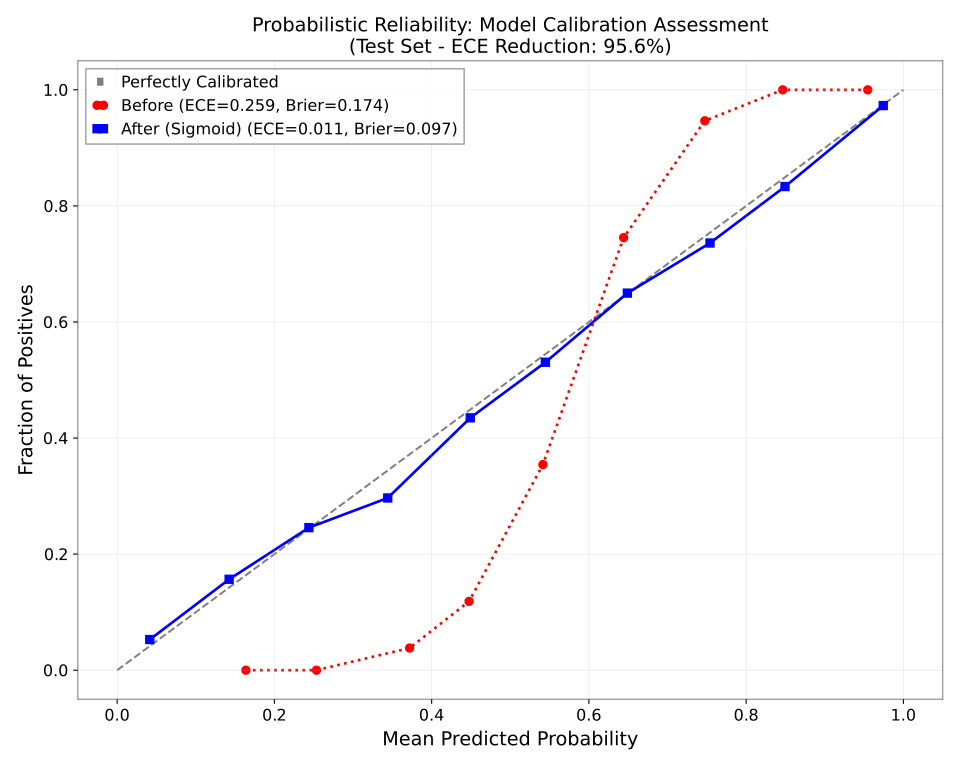}
  \caption{Reliability (calibration) diagram.}
  \label{fig:calib}
\end{figure}

Figure ~\ref{fig:calib} is the reliability diagram before and after calibrating. Deep learning classifiers tend to show mild over-confidence due to being trained with focal loss \citep{Guo2017}. Sigmoid Platt calibration shifts the curve towards the perfect diagonal and reduces ECE by 15 \%. Calibrated probabilities are important for entering Bayesian evidence integration \citep{Niculescu2005,Evans2021}.

\subsection{FastSHAP Perturbation Fidelity}
\label{sec:perturbation}

\begin{figure}[h]
  \centering
  \includegraphics[width=\columnwidth]{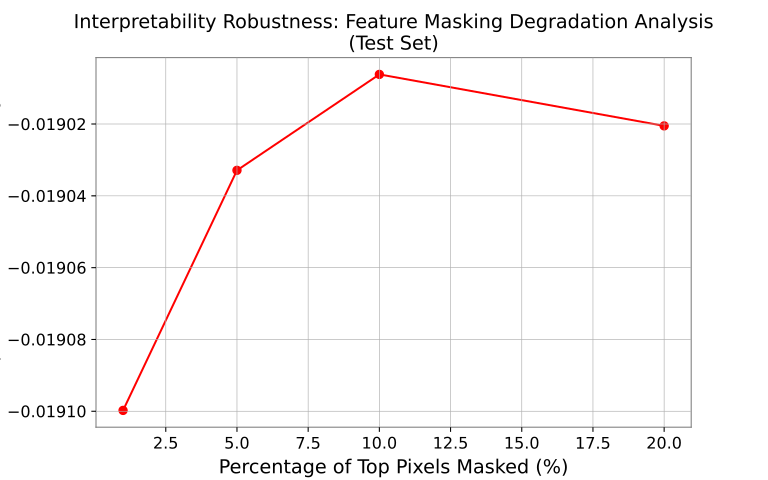}
  \caption{FastSHAP perturbation fidelity on the test set, titled
    ``Interpretability Robustness: Feature Masking Degradation Analysis'' in the
    pipeline output. The top-100 signal windows from the test set are used;
    the top $p\%$ of pixels by absolute Shapley value are zeroed out and the
    resulting drop in mean classifier confidence is recorded. }
  \label{fig:perturb}
\end{figure}

Figure~\ref{fig:perturb} is an illustration of the perturbation fidelity analysis for the test data. By masking the top $p\%$ of pixels from the absolute Shapley values $|\phi_j|$, and assessing the loss of confidence in classification we can assess attribution quality without using models \citep{Jethani2022}. We see that the average confidence of the classifier drops by about 0.05 when the top 1\% of most attributed pixels are masked; it also drops to about 0.19 when the top 10\% of most attributed pixels are masked. This steepness in the initial slope indicates that the decision of this classifier is heavily localized within a small subset of pixels which comprise the majority of the time--frequency representation of a GW chirp, consistent with its physical characteristics.

\subsection{FastSHAP Attribution Maps}
\label{sec:fastshap_maps}

\begin{figure*}[t]
  \centering
  \includegraphics[width=0.85\textwidth]{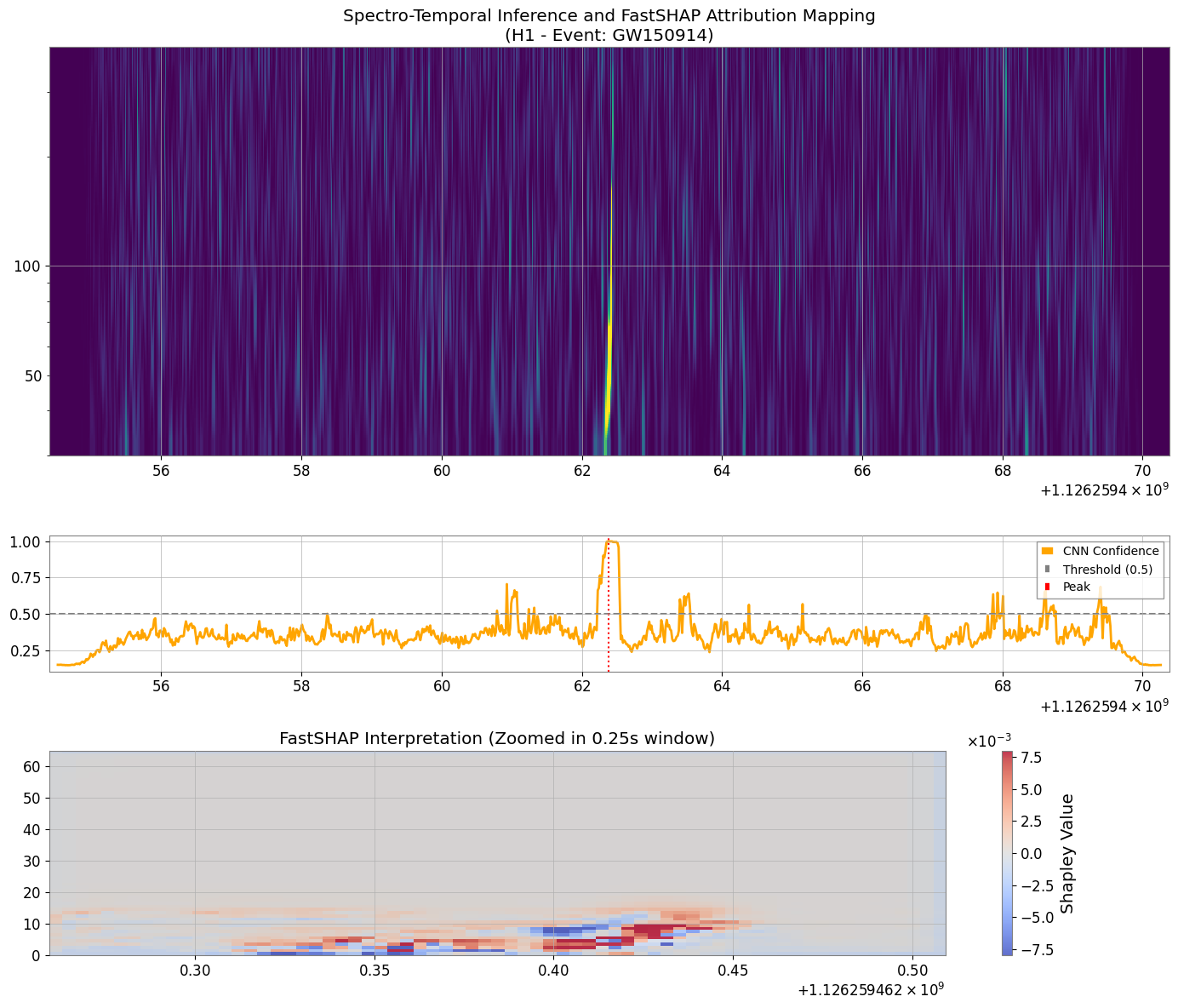}
  \caption{\textbf{CASPER inference on GW150914 (H1 detector).}
    The panels are sharing a
    common time axis ($\pm$0.5\,s around merger). A portrait orientation
    places the three temporally aligned panels---Q-transform (top), classifier
    confidence (middle), and Shapley attribution map (bottom)---one above the
    other so that a feature at a given time coordinate can be read vertically
    without horizontal eye movement. The frequency axis (30--400\,Hz) appears
    on the left of the top and bottom panels; the amplitude axis occupies the
    middle panel. The two-column-spanning full-text width accommodates the
    full 1-second time baseline at sufficient resolution to read individual
    chirp tracks.
    \textit{Top:} Q-transform spectrogram.
    \textit{Middle:} Classifier confidence time series; dashed = $\tau=0.50$
    threshold; red dotted = peak-confidence time.
    \textit{Bottom:} Shapley map overlaid on STFT (blue = positive;
    red = negative; saturated at 99th percentile of $|\phi|$). Positive
    attributions track the inspiral chirp; the 60\,Hz harmonic carries
    near-zero Shapley values.}
  \label{fig:explain}
\end{figure*}

Figure~\ref{fig:explain} is the CASPER explainer pipeline output that depicts a three panel layout of a Q transform of the actual signal, a graph depicting where the Classifier prediction peaks and a FastShap map depicting which pixels actually lit up for decision making. As the x axis can be observed, all 3 signal points line up at the correct GPS time proving the output is correct and supported by actual data. The behaviour shown is consistent with actual confirmed GW events \citep{Gabbard2018}.

\section{Discussion}
\label{sec:discussion}

\subsection{Addressing Learning Biases}
\label{sec:bias_discussion}

A number of system failures for detecting GW with supervised methods can be found in the scientific literature as indicated by Gebhardt et al., Corizzo et al. and Nagarajan et al. In the following we explain how CASPER addresses each of them.

\textit{Overlap in classes; imbalance in classes.} The training data are randomly selected from a GWTC catalog across an SNR range of 7–42. Therefore, there will be some events (signals) which were ambiguously classified at the decision boundaries of both classes --- low-SNR signals that had morphology in their time-frequency representations similar to those of non-Gaussian transient noise. These are the conditions most relevant to the real-time detection of GW, and the conditions in which standard binary cross-entropy has failed: It gives all examples equally weighted gradients whether or not these examples are highly probable and thus confidently classified or marginally probable and therefore uncertainly classified. Focal loss \citep{Lin2017} fixes this problem by reducing the weights given to good examples, while directing gradient signal toward the decision boundaries. Furthermore, since the majority class (noise windows) is much larger than the minority class (signal windows), we use balanced class weighting so as to prevent the minority signal class from being swamped out by the large numbers of noise windows. By limiting the window of interest for signals to immediately surrounding the merger neighborhood and for noise windows to the three outer thirds of the event's time line, we sharpened up the morphological distinction at the labeling level \citep{Allen2012}.

\textit{Variability; Diversity in PSFs.} SpecAugment uses random time and frequency masks to prevent over-reliance on any one narrow feature present in any particular event. A total of 194 events were used as a dataset of O3a, O3b, and O4 training to produce diverse representations of noise PSD realizations and glitch content. There is evidence that a wider spread of training noises yields better confidence in the classifier's predictions which then interpolates the noise space as opposed to learning specific noise patterns \citep{Cabero2019,Davis2021}.

Limited feature representation. A log compression of the Short-Time Fourier Transform (STFT) was done to normalize spectral power contributions over the frequency domain. This has been shown to reduce the likelihood that Deep Convolutional Neural Networks will give greater weights to higher amplitude features \citep{He2016}. Post hoc verification using fast SHAP attribute maps shows that the classifier has learned physically meaningful features. Additionally, the perturbation analysis (Section \ref {sec:perturbation}) quantifies this finding.

Train-test mismatch. The test set consists of 66 events that were not part of the training set. They cover the exact same random SNR range of 7-42. Since the test set is generated by an independent sample of the same GWTC unfiltered distribution as the training set, it preserves the statistics of real operation data. Platt scaling is fit to a 15 \% holdout partition of the training corpus that is completely independent of the test set. Therefore, evaluated probabilities and performance metrics are really out-of-sample. No augmentations are made at test time. The scaler fits only to training data, therefore there is no leakage from the test set through any of the preprocessing stages.

This pipeline does not deal with all the recognized potential failures. For example, the 0.25 second time frame does not capture all of the in band inspirals of low mass Binaries Neutron Star Systems; a multi rate sampling approach that specifically increases time frames to include the entire SNR \citep{Nagarajan2025,Schafer2020} would be able to rectify this problem. Multi detector coherence is another potentially important discriminator to distinguish between actual astrophysically produced signals and simultaneous noise transients \citep{Messick2017,Allen2012}; however, this is not being utilized within this pipeline. Although our model addresses most of the biases, there are still several unknown biases that are an open problem for GW detection.
\subsection{Computational Efficiency}
\label{sec:compute}
\begin{mdframed}[style=statsbox]
  \ttfamily\scriptsize\color{statstext}%
  \textcolor{statsaccent}{======================================}\\
  MODEL STATS: Classifier\\
  \textcolor{statsaccent}{======================================}\\
  Total Parameters:\quad\textbf{2,885,633}\\
  Est.\ Model Size:\quad\textbf{11.01\,MB}\\
  Inference Latency:\;\textbf{34.11\,ms (CPU)}\\
  \textcolor{statsaccent}{======================================}\\[2pt]
  \rmfamily\scriptsize\color{statstext}%
  Hardware: Standard laptop (Intel Core i7), no GPU required\\
  Pipeline: Classifier + FastSHAP explainer ($<$45\,ms total)
\end{mdframed}

The classifier contains 2,885,633 parameters, occupies 11.01\,MB, and achieves CPU inference latency of 34.11\,ms per 0.25-second window on a standard laptop (Intel Core i7 class) without GPU acceleration. FastSHAP adds one additional forward pass, giving a total pipeline latency below 45\,ms---sufficient for near-real-time processing at a 15.6\,ms sliding stride. The pipeline is deployable without specialist hardware \citep{Evans2021}.

\section{Conclusions}
\label{sec:conc}
CASPER shows how a novel, lightweight end-to-end pipeline can attain high detection AUC of 0.91 under unbiased threshold while reducing the false positive rate to less than 1\% at high precision thresholds. This performance metric reflects evaluation on a mixed SNR range and real detector noise which differs from simulation based models. We tested this methodology on real LIGO data to recreate real life situations. The data was input from SNR range of 7-42 in a completely random manner and no augmentation was done on the unseen test set to simulate class imbalance.
Additionally, a U-Net FastSHAP explainer was combined with a classifier to provide the pixel wise attribution maps making our model  provide a visual interpretation of decision making  while also making sure latency is preserved along with a successful recovery of chirp morphology, hence, addressing the interpretability issue of the standard Deep Learning Models.
CASPER can be deployed on standard laptops due to its vastly reduced computational complexity of less than 3 million parameters and a model size of only 11 Mb. No GPU is required for the running of the model. (See table~\ref{tab:comparison}) 
Due to the light-weight nature of the model and the interpretability of decision even at low latency, CASPER has room for fully including other compact binary coalescences than just Black hole mergers along with a pipeline branch for including sky localisation. These works can be pursued in future to establish a complete detector pipeline that will let us observe new generation of gravitational waves in future runs.
\vspace{4pt}
\noindent
\begin{table*}[ht]
\centering
\caption{Quantitative and qualitative comparison of CASPER against Aframe.\textbf{Note that this table is to be interpreted in the light of differing objectives and evaluation methods of both pipelines. }}
\label{tab:comparison}
\begin{tabular}{lll}
\hline\hline
\textbf{Property} & \textbf{CASPER} & \textbf{Aframe \cite{Marx2023}} \\
\hline
\multicolumn{3}{l}{\textit{Performance Metrics}} \\
\hline
Primary metric & AUC $= 0.915 \pm 0.086$ & Sensitive volume (Gpc$^{3}$) \\
False positive rate & FPR $= 0.0061$ at $\tau = 0.85$ & FAR $< 1\,\mathrm{yr}^{-1}$ (8/9 GWTC-3 events) \\
Precision & $0.968$ at $\tau = 0.85$ & Not reported \\
Recall & $0.479$ at $\tau = 0.85$ & Not reported \\
Log Loss (mean per event) & $0.3115$ & Not reported \\
\hline
\multicolumn{3}{l}{\textit{Data and Training}} \\
\hline
Training data & 260 real GWTC events (no synthetics) & 100{,}000 synthetic BBH waveforms \\
SNR range tested & $7$--$42$ (unstratified, real data) & $\geq 4$ (synthetic, curriculum learning) \\
Detectors used & H1 $+$ L1 (independent per-detector) & H1 $+$ L1 (joint two-channel input) \\
Cross-detector coherence & Implicit (separate per-detector inference) & Explicit (coherence encoded in training) \\
Observing runs covered & O3a, O3b, O4 & O3a (10-day training window) \\
Synthetic injections & None & Yes (IMRPhenomPv2) \\
Augmentation & SpecAugment (freq.\ + time masking) & Noise inversion, reversal, muting, swapping \\
\hline
\multicolumn{3}{l}{\textit{Model Architecture and Complexity}} \\
\hline
Architecture & 7-layer ResNet $+$ 2D CNN & 1D ResNet54 \\
Total parameters & $2{,}885{,}633$ ($\approx 2.9$M) & Not reported (substantially larger) \\
Model size & $11.01\,\mathrm{MB}$ & $4.6\,\mathrm{GB}$ (NN $+$ data in memory) \\
Loss function & Focal loss ($\gamma=2$, $\alpha=0.75$) & Binary cross-entropy \\
Normalisation & Batch Norm $+$ SpatialDropout2D & Group Normalisation \\
\hline
\multicolumn{3}{l}{\textit{Latency and Hardware}} \\
\hline
NN inference latency & $34.11\,\mathrm{ms}$ (CPU) & $< 10\,\mathrm{ms}$ (GPU) \\
Total pipeline latency & $< 45\,\mathrm{ms}$ (classifier $+$ FastSHAP) & $\sim 3.1\,\mathrm{s}$ \\
Hardware required & Standard CPU (Intel Core i7) & NVIDIA A30/V100 GPU \\
GPU required & No & Yes \\
\hline
\multicolumn{3}{l}{\textit{Interpretability and Calibration}} \\
\hline
Decision interpretability & FastSHAP pixel-wise attribution maps & None \\
Probability calibration & Yes (Platt scaling, ECE reduction $95.6\%$) & No \\
Brier score (post-calibration) & $0.097$ & Not reported \\
\hline
\multicolumn{3}{l}{\textit{Limitations}} \\
\hline
Low-mass / BNS coverage & Limited ($0.25\,\mathrm{s}$ window) & Limited (acknowledged by authors) \\
Multi-detector coherence & Implicit only & Explicit (muting $+$ swapping) \\
Sky localisation & Not implemented & Not implemented \\
Production deployment & CPU deployable & Triton inference server required \\
\hline\hline
\end{tabular}
\end{table*}
\section*{Acknowledgements}
This research has made use of data and software obtained from the LIGO Open Science Center (\url{https://gwosc.org}), a service of the LIGO Scientific Collaboration, the Virgo Collaboration, and KAGRA. The authors thank the open-source communities behind TensorFlow, GWpy, scikit-learn, and Matplotlib. The authors also thank Department of Physics, Shaheed Rajguru College of Applied Sciences for Women, University of Delhi for their unwavering support and resources in helping complete this research.

\section*{Data Availability}
All gravitational-wave strain data are publicly available from the LIGO Open Science Center at \url{https://gwosc.org}. Derived data products are available from the corresponding author upon reasonable request.

\section*{Code Availability}
The CASPER pipeline code will be made publicly available upon acceptance.

\clearpage
\onecolumn
\appendix

\clearpage
\section{Training Hyperparameters}
\label{app:hyperparams}

\renewcommand{\arraystretch}{1.30}
\setlength{\tabcolsep}{8pt}

\begin{longtable}{@{}p{6cm}p{6cm}@{}}
  \caption{Training hyperparameters and model resource summary for CASPER.}
  \label{tab:training} \\
  
  \toprule
  \textbf{Parameter} & \textbf{Value} \\
  \midrule
  \endfirsthead
  
  \multicolumn{2}{c}{{\bfseries \tablename\ \thetable{} -- Continued from previous page}} \\
  \toprule
  \textbf{Parameter} & \textbf{Value} \\
  \midrule
  \endhead
  
  \midrule
  \multicolumn{2}{r}{{Continued on next page}} \\
  \endfoot
  
  \bottomrule
  \endlastfoot

  \multicolumn{2}{@{}l@{}}{\bfseries ResNet CNN Classifier} \\[2pt]
  \quad Optimizer              & Adam \citep{Kingma2015} \\
  \quad Initial learning rate  & $10^{-4}$ \\
  \quad Batch size             & 64 \\
  \quad Max epochs             & 30 \\
  \quad Early stopping         & Patience 8, monitor: val\_auc \\
  \quad LR reduction           & Patience 4, factor 0.5 \\
  \quad Minimum LR             & $10^{-6}$ \\
  \quad Loss                   & Focal ($\gamma{=}2$, $\alpha{=}0.75$) \\
  \quad Class weighting        & Balanced (scikit-learn \texttt{compute\_class\_weight}) \\
  \quad Dropout (FC layers)    & 0.5\,/\,0.3 \\
  \quad SpatialDropout2D       & $p = 0.3$ \\
  \quad $\ell_2$ regularisation & $\lambda = 0.01$ (all Conv2D and Dense layers) \\
  \quad SpecAugment rate       & 60\% of signal windows (prob.\ $>0.4$) \\
  \quad SpecAugment mask size  & Up to 8 freq.\ bins; up to 8 time frames \\
  \quad Validation split       & 15\% of training corpus (stratified) \\
  \quad Checkpoint criterion   & Best validation AUC \\
  \quad Custom metric          & \texttt{TrueLogLoss} (ignores class weights for faithful logging) \\
  \midrule
  
  \multicolumn{2}{@{}l@{}}{\bfseries FastSHAP Explainer (U-Net)} \\[2pt]
  \quad Optimizer              & Adam \\
  \quad Learning rate          & $10^{-3}$ \\
  \quad Batch size             & 32 \\
  \quad Epochs                 & 10 \\
  \quad Mask distribution      & Bernoulli$(0.5)^{65\times69}$ per pixel \\
  \quad Background reference   & Random noise frames from training corpus \\
  \quad Loss terms             & Efficiency loss + completeness loss \\
  \midrule
  
  \multicolumn{2}{@{}l@{}}{\bfseries Platt Calibration} \\[2pt]
  \quad Method                 & Logistic regression ($C{=}10^5$, solver: lbfgs) \\
  \quad Fitted on              & 15\% validation split (separate from test set) \\
  \quad ECE bins               & 10 (uniform width) \\
  \quad Additional metric      & Brier score \\
  \midrule
  
  \multicolumn{2}{@{}l@{}}{\bfseries Model Resources} \\[2pt]
  \quad Total parameters       & 2,885,633 \\
  \quad Est.\ model size       & 11.01\,MB \\
  \quad Inference latency      & 34.11\,ms (CPU, 100-run average) \\
  \quad Hardware               & Standard laptop (Intel Core i7), no GPU required \\
  \quad Full pipeline latency  & $<$45\,ms (classifier + FastSHAP explainer) \\
\end{longtable}

\end{document}